\documentclass[aps,12pt,onecolumn,superscriptaddress,showpacs,floatfix]{revtex4}

\usepackage[dvips]{epsfig}
\usepackage[dvips]{color}

\newcommand{\gdtio}{Gd$_2$Ti$_2$O$_7${}}
\newcommand{\gdsno}{Gd$_2$Sn$_2$O$_7${}}
\newcommand{\JPCM}{J.Phys.:Condens.Matter}
\newcommand{\PR}{Phys. Rev.}
\newcommand{\figwidth}{0.9\columnwidth}

\begin{document}

\title{Observation of a transverse magnetization in the
ordered phases of the pyrochlore magnet \gdtio.}

\author{V.N. Glazkov}

\affiliation{Commissariat \`a l'Energie Atomique, DSM/DRFMC/SPSMS,
38054 Grenoble, Cedex 9, France}

\affiliation{P.~L.~Kapitza Institute for Physical Problems RAS,
119334  Moscow, Russia}

\author{C. Marin and J.-P. Sanchez}

\affiliation{Commissariat \`a l'Energie Atomique, DSM/DRFMC/SPSMS,
38054 Grenoble, Cedex 9, France}

\date{\today}

\begin{abstract}
We have performed a detailed transverse magnetization study of the
pyrochlore antiferromagnet \gdtio{}. A transverse magnetization of
about $10^{-3}M_{sat}$ is observed in the
low-temperature ordered phases. These measurements result in the
refinement of the \gdtio{} phase diagrams. Observation of a
transverse magnetization indicates loss of the cubic symmetry in
some of the magnetic phases and provides new information for a
better understanding of the complicated magnetic ordering of
\gdtio{}.
\end{abstract}


\maketitle

\vspace*{5mm}

Gadolinium compounds  \gdtio{} and \gdsno{} provide good examples
of the Heisenberg antiferromagnets on the pyrochlore lattice. The
magnetic Gd$^{3+}$ ions are located at the vertices of the network
of corner sharing tetrahedra. The nearest neighbor Heisenberg
exchange interaction is strongly frustrated in this geometry ---
nearest neighbor exchange coupling alone is not sufficient to
select a unique ground state of the pyrochlore antiferromagnet.
However, both of these compounds orders near 1K \cite{sosin}. This
ordering is caused by the subtle interplay of weaker interactions
(e.g. dipolar coupling, further neighbor exchange couplings and
single-ion anisotropy).

The magnetic phase diagrams of \gdtio{} \cite{Petrenko}, as
determined from the specific heat measurements, reveal a variety
of ordered magnetic phases. Two phase transitions are observed in
zero external field at T$_{N1}$=1.05K and T$_{N2}$=0.75K. At low
temperatures a transition to the saturated phase is observed at a
field near 6T, while an additional phase transition is found near
H$\sim$3T for the $\mathbf{H}||[110],[111]$. The later phase
transition is reported to be absent for the $\mathbf{H}||[211]$.
The two-step phase transition in zero field and  the field-induced
transitions in this compound are not understood yet. Moreover, the
experimentally determined phase diagram for the
$\mathbf{H}||[211]$ \cite{Petrenko} is not consistent with the
general thermodynamic restrictions \cite{yip} forbidding existence
of a point where three second-order transition lines meet.
Identification of the zero-field magnetic structure by the means
of  neutron scattering \cite{gdtio-structure} yields a
4$\mathbf{k}$-structure with $\mathbf{k}=(1/2,1/2,1/2)$ and where
1/4 of the Gd ions remains disordered at T$_{N2}<T<T_{N1}$. These
ions order  at the second phase transition at T$_{N2}$.  The
high-temperature ordered phase ($T_{N2}<T<T_{N1}$) preserves cubic
symmetry \cite{symmetry-comment}, while the symmetry of the
low-temperature phase is not clearly understood being hidden in
the complications of the 4$\mathbf{k}$-structure.

\begin{figure}
  \centering
  \epsfig{file=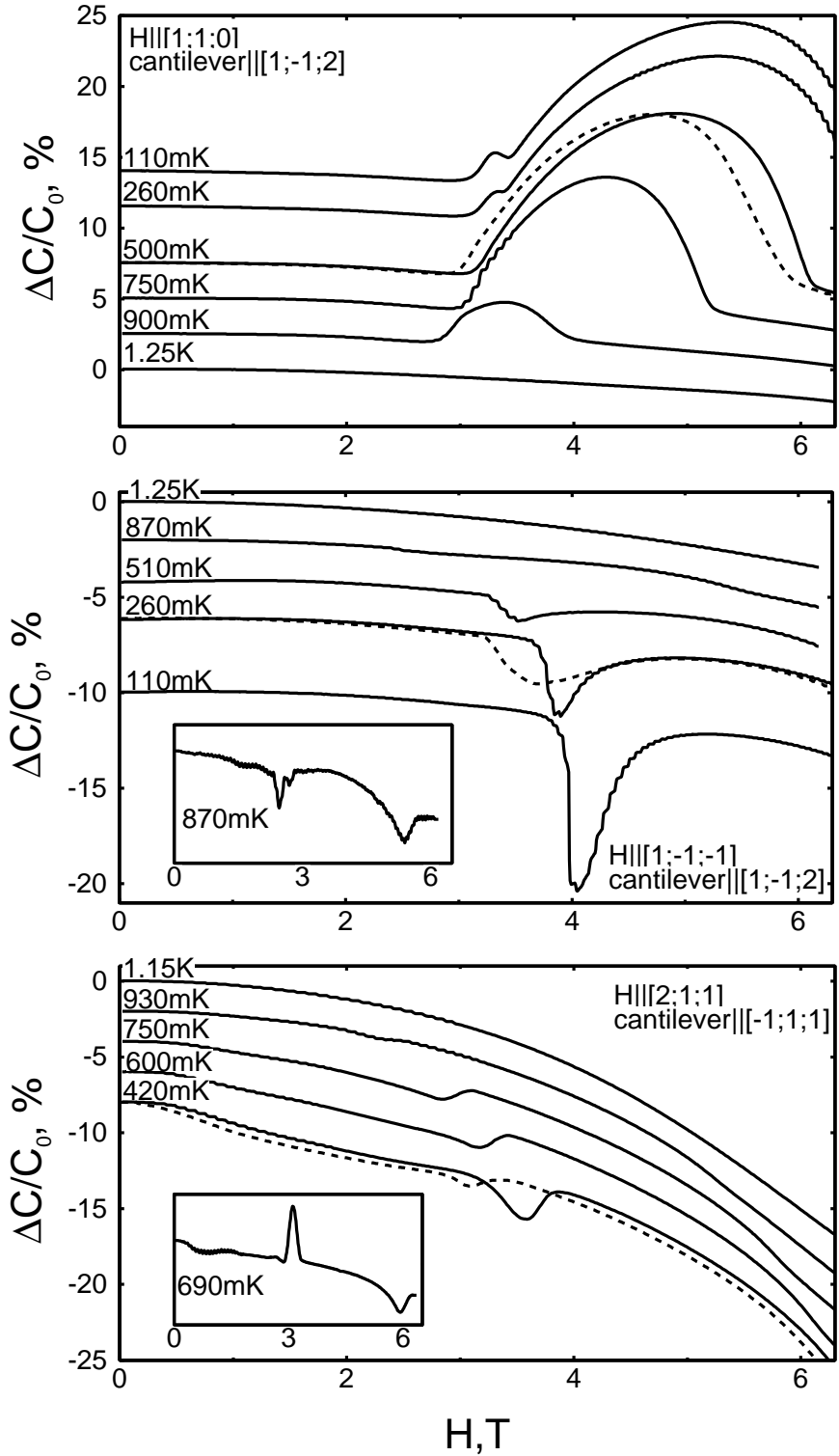, width=0.7\columnwidth, clip=}%fig1-hor.eps or fig1.eps
  \caption{Field dependences of the torquemeter response for different
  field orientations. Inserts:   field dependences of the field derivative of the
  torquemeter response for the corresponding orientations.
  Solid curves (main panels and inserts): data measured on
  increasing magnetic field. Dashed curves: examples of the data
  measured on decreasing field.}\label{fig:data}
\end{figure}

The present work was stimulated by the prediction of a possible
appearance of a transverse magnetization in the classical
Heisenberg pyrochlores above some critical fields
\cite{glazkov-gdti}. A transverse magnetization change is a
sensitive indicator for the magnetic phase transition
\cite{abarzhibazhan}. The transverse magnetization could  appear
in the crystal if the magnetic field direction deviates from the
main axes of the susceptibility tensor. In the case of the cubic
symmetry the susceptibility tensor is isotropic. Thus, no
transverse magnetization could be observed in the magnetic phases
with cubic symmetry.

To measure  the transverse magnetization we have used a
capacitance torquemeter. The sensitive element of the torquemeter
was a plane capacitor with the  parallel plates formed by the
 rigid base and the flexible bronze cantilever.
The sample was glued to the flexible cantilever and the
experimental cell was mounted on the dilution refrigerator
equipped with a 6T cryomagnet. The cantilever was thermalized with
the mixing chamber. The magnetic field was applied perpendicular
to the capacitor plates. The experimental cell was connected to
the General Radio capacitance bridge, which was balanced before
the measurements.  The imbalance of the bridge was measured as a
function of the applied magnetic field. Maximal capacitance change
during the measurements was about 20\%. Capacitance of the
experimental cell at H=0 was about 1.4pF. Cantilever deformations
were found to be always reversible. A \gdtio{} single crystal
was prepared following the same technique as described in
\cite{gdtio-muons}. Samples were oriented and cut in rectangular
shape with the dimensions of about $0.6\times0.9\times1.3$mm$^3$.

The ideal torquemeter response is only due to the transverse
magnetization component $\mathbf{M}_{\perp}$ aligned along the
flexible cantilever. As the magnetic field $\mathbf{H}$ is applied
perpendicular to the cantilever plate, the torque
$\mathbf{M}_{\perp}\times\mathbf{H}$ acts on the sample. This
torque is compensated by the bending of the cantilever resulting
in the change of the experimental cell capacitance $\Delta C$
detected as the bridge imbalance. Thus, the torquemeter output
could be expressed as $U(T,H)\propto\Delta C\propto M_{\perp}H$.

A real torquemeter is also slightly sensitive to the longitudinal
magnetization component  $\mathbf{M}_{\parallel}$ via the
magnetic-balance response due to the small uncontrolled field
gradient at the sample position. Another uncontrolled effect is a
torque  due to the demagnetization tensor anisotropy for the
non-spherical samples --- an elongated sample in a magnetic
field tends to rotate its longest axis parallel to the field
direction. These two parasitic effects should be present even for
the cubic crystals. They are proportional to the $H^2$ at low
fields and are expected to change their field dependence at the
saturation field. They are expected to be almost temperature
independent since the magnetic susceptibility of \gdtio{} does not
change strongly below 1K \cite{sosin}.

\begin{figure}
  \centering
  \epsfig{file=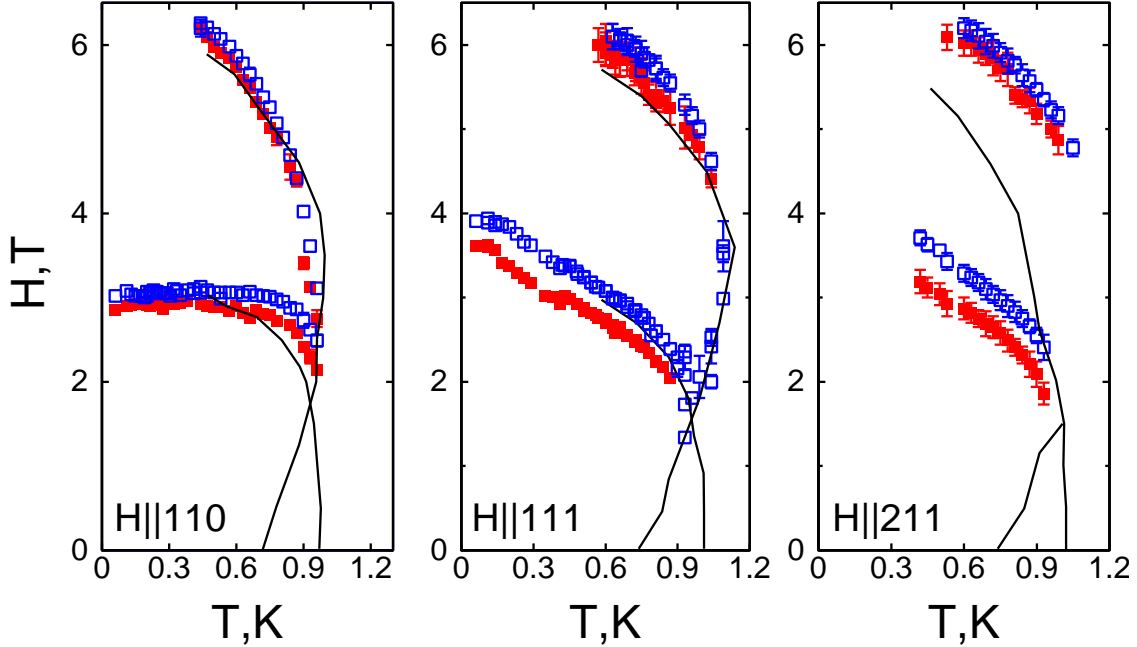, width=\figwidth, clip=}
  \caption{(color)H - T phase diagrams. Open symbols (blue): data taken on increasing field.
  Closed symbols (red): data taken on
  decreasing field. Solid curves: data from Ref.\cite{Petrenko}}\label{fig:ht}
\end{figure}

\begin{figure}
  \centering
  \epsfig{file=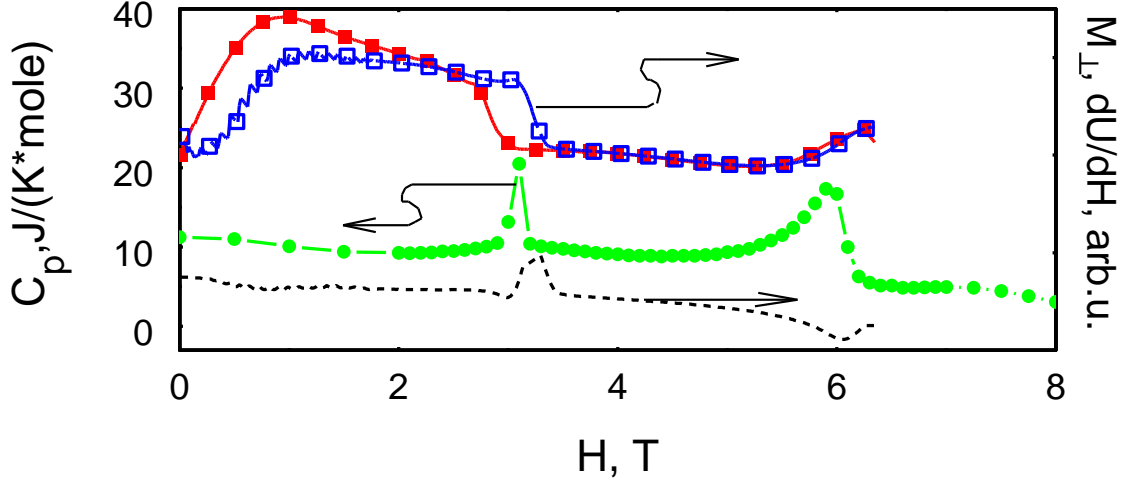, width=\figwidth, clip=}
  \caption{(color)Field dependence of the specific heat (circles, green) and its correspondence with the results of
  transverse magnetization measurements. Dashed line --- field derivative of the torquemeter output, squares ---
  determined field dependences of the transverse magnetization measured at increasing
  (open, blue) and decreasing  (closed, red) field. For all curves T=620mK, $\mathbf{H}||[211]$.}\label{fig:c112}
\end{figure}

Field dependences of the torquemeter response taken at different
temperatures are shown in Figure \ref{fig:data}. Above 1K the
occurrence of a transverse magnetization is forbidden by the cubic
symmetry of the paramagnetic phase. The observed response is due
to the parasitic effects of longitudinal magnetization and
demagnetization-induced torque. The maximal amplitude of the
response due to these parasitic effects can be estimated from the
amplitude of the torquemeter response at 6T in the paramagnetic
phase. The shape of the torquemeter response changes as the
temperature decreases below 1K, i.e, when the sample enters into
the ordered state. For the $\mathbf{H}||[110],[111]$ a strong
hump-shaped deviation of the torquemeter output from the
high-temperature behavior appears near 3T. The amplitude of this
hump at low temperatures exceeds significantly the response that
could be ascribed to the parasitic effects, i.e., observed
hump-shaped signals are due to the appearance of the transverse
magnetization. At high fields the hump disappears and the
torquemeter response is close to that observed in the paramagnetic
state. For the $\mathbf{H}||[110]$ this hump has  well defined
low-field and high-field edges, which allow  to estimate the
critical fields of appearance and disappearance of the transverse
magnetization. For the  $\mathbf{H}||[111]$ the transverse
magnetization signal has a sharp low-field edge only.

Evolution of the torquemeter response for the $\mathbf{H}||[211]$
geometry is different. As the sample is cooled down below 1K the
\emph{low-field} part of the torquemeter response deviates from
the paramagnetic $H^2$ behavior. Developing hump demonstrates a
sharp high-field edge. For this orientation we were unable to
perform measurements below 400mK for technical reasons. Amplitudes
of the torquemeter response deviations from the high-temperature
behavior is not high enough to ascribe unambiguously these changes
to the transverse magnetization effects. However, to attribute
this signal to the parasitic effects, one has to assume that the
sample magnetization at low fields (below 3T) almost reaches the
saturation and then decreases. Neglecting this unlikely scenario,
we conclude that this response is also connected with the
appearance of the transverse magnetization.

A transition to the saturated phase can be found for the
$\mathbf{H}||[111],[211]$ orientations by taking the field
derivatives of the torquemeter output (see inserts in Figure
\ref{fig:data}). They change sharply near 6T  from the
field-dependent behavior to a constant owing to the change of the
field dependences of the parasitic effects at the saturation
field.

A hysteresis both in the position and in the form of the
transverse magnetization related hump is observed for all
orientations (see Figure \ref{fig:data}). At increasing fields
appearance and disappearance of the transverse magnetization are
always observed at higher field values than for decreasing field.
The hump is also more pronounced on the experimental curves taken
on increasing field. This hysteresis does not disappear as the
magnetic field sweep rate is tenfold decreased.

\begin{figure}
  \centering
  \epsfig{file=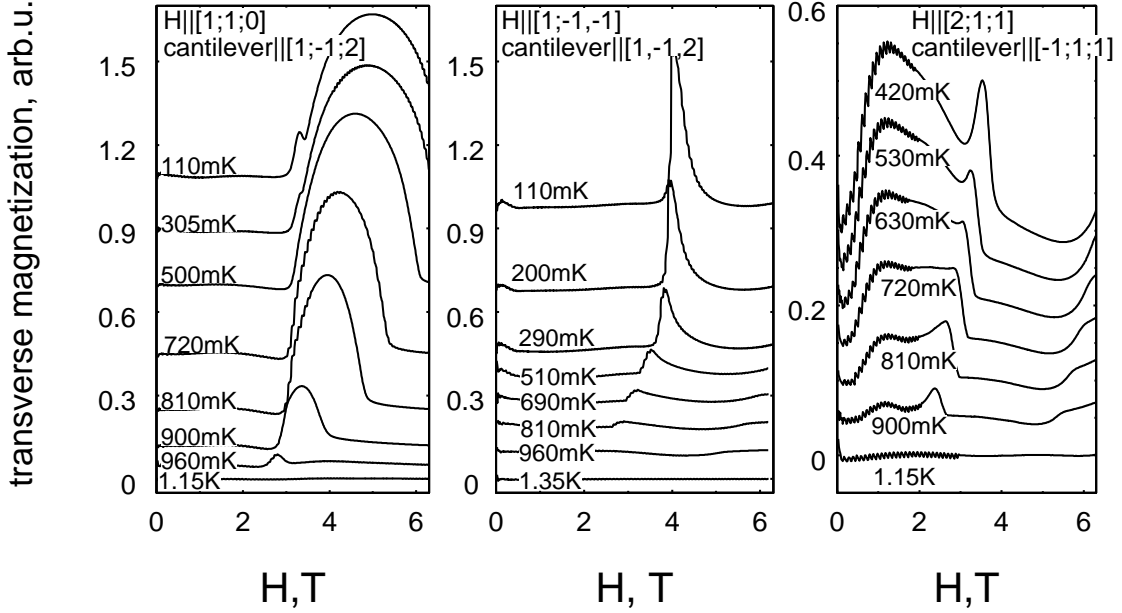, width=\figwidth, clip=}
  \caption{Field dependences of the normalized transverse magnetization (see text).
  All curves correspond to the measurements done on increasing magnetic field.}\label{fig:m}
\end{figure}

Identified features of the torquemeter response (i.e., fields of
appearance and disappearance of the transverse magnetization and
fields of the transition to the saturated phase) were used to draw
the $(H-T)$ diagrams shown in Figure \ref{fig:ht}. These features
correlate very well with the known phase diagrams \cite{Petrenko}
for the $\mathbf{H}||[110],[111]$ orientations of the applied
magnetic field. However, our results for the $\mathbf{H}||[211]$
suggest that there is a phase transition line near H=3T that was
not detected in the known specific heat measurements
\cite{Petrenko}. This suggested phase transition line is in
agreement with general thermodynamic restrictions \cite{yip}. We
have measured the field dependence of the specific heat in the
$\mathbf{H}||[211]$ geometry at $T=620$mK using a Quantum Design
PPMS calorimeter (Figure \ref{fig:c112}). Our specific heat data
demonstrates clear transitions near 3.1T and 6.1T in perfect
agreement with the transverse magnetisation measurements.

Finally, we can recover the field dependences of the transverse
magnetization. To account for the parasitic effects contribution,
we subtracted the torquemeter response measured in the
paramagnetic phase (at $T_0=$1.15\ldots1.35K) from the low
temperature response. This approach yields the following
expression for the transverse magnetization per unit mass:

\begin{equation}\label{eqn:m-recovery}
  |M_{\perp}(T,H)|\propto\frac{\left|U(T,H)-U(T_0,H)\right|}{Hm}
\end{equation}

\noindent here $m$ is the sample mass. The field dependences of
the so extracted transverse magnetization are presented in Figure
\ref{fig:m}.  Note, that for the $\mathbf{H}||[211]$ orientation
at $T_{N1}>T>T_{N2}$ (810mK and 900mK curves) the transverse
magnetization appears only above a given magnetic field. This
observation is in agreement with the high symmetry of the
high-temperature ordered phase, which forbids the transverse
magnetisation. The critical field for the appearance of the transverse
magnetization corresponds well to the known phase boundary between
ordered phases.

The magnitude of the transverse magnetization can be roughly
estimated from the elastic constant of the cantilever. This yields
a value of $3\cdot10^{-3}M_{sat}$ for the maximal amplitude of the observed transverse magnetization.

Summarizing our results, we have observed  a transverse
magnetization  in the low-temperature magnetically ordered phases
of the pyrochlore antiferromagnet \gdtio{}.  This means
unambiguously that the cubic symmetry is lost in the
low-temperature ordered phases of \gdtio{}. We have also refined the
magnetic phase diagrams and have detected a new phase transition
line for $\mathbf{H}||[211]$.

\acknowledgements

The authors thank M.Zhitomirsky (CEA-Grenoble/DRFMC/SPSMS) for his
continuous interest to this work and numerous fruitful and
stimulating discussions, I.Sheikin (CNRS-Grenoble/HMFL) and
C.Marcenat (CEA-Grenoble/DRFMC/SPSMS) for discussions concerning
experimental techniques, S.S.Sosin  and A.I.Smirnov (Kapitza
Institute) for helpful discussions on the interpretation of the
experimental results.


\vspace{5mm}

\begin{thebibliography}{50}
\bibitem{sosin} Bonville P, Hodges J A, Ocio M, Sanchez J P, Vulliet P,
Sosin S and Braithwaite D 2003  \JPCM{} \textbf{15} 7777



\bibitem{Petrenko} Petrenko O A, Lees M R, Balakrishnan G, and McK Paul D 2004,
\PR{} B \textbf{70} 012402


\bibitem{yip} Yip S K, Li T and Kumar P 1991  \PR{} B \textbf{43} 2742

\bibitem{gdtio-structure} Stewart J R , Ehlers G, Wills A S,
Bramwell S T and Gardner J S 2004 \JPCM{} \textbf{16} L321

\bibitem{symmetry-comment} As one can see from
Ref.\cite{gdtio-structure}, at $T_{N1}>T>T_{N2}$ all
$\langle111\rangle$ directions are the 3-rd order symmetry axes.
Thus, the symmetry of this phase is at least tetrahedral and belongs
to the cubic symmetry class.


\bibitem{glazkov-gdti} Glazkov V N, Zhitomirsky M E,
Smirnov A I, Krug von Nidda H A, Loidl A, Marin C, and Sanchez J P
2005 \PR{} B \textbf{72}  020409(R)

\bibitem{abarzhibazhan} Abarzhi S I, Bazhan A N, Prozorova L A and
Zaliznyak I A 1992 \JPCM{} \textbf{4}  3307

%\bibitem{sosin-jetp} Zhitomirskii M E, Petrenko O A, Petrov S V, Prozorova L A ,
%Sosin S S 1995 Sov.Phys. JETP \textbf{81} 185

\bibitem{gdtio-muons} Yaouanc A, Dalmas de R\'{e}otier P, Glazkov V, Marin C,
Bonville P, Hodges J A, Gubbens P C M, Sakarya S, and Baines C 2005
Phys. Rev. Lett. \textbf{95} 047203

\end{thebibliography}
\end{document}